# Human diversity fuels collective creativity that large language models cannot simulate or sustain

Mengchen Dong†, Hiromu Yakura†

Center for Humans and Machines, Max Planck Institute for Human Development, Berlin, Germany

## Author notes

Both authors contributed equally to this work.

Correspondence concerning this article should be addressed to Mengchen Dong, Max Planck Institute for Human Development, Lentzeallee 94, 14195 Berlin, Germany.
Email: dong@mpib-berlin.mpg.de

The authors used a large language model (Claude, Anthropic) to assist with language editing, refining the logical flow of arguments, and drafting analysis and visualization code. The study design, data collection, analyses, and results are entirely the authors' work. The authors reviewed, verified, and edited all text and take full responsibility for the content of the manuscript.

# Human diversity fuels collective creativity that large language models cannot simulate or sustain

## Abstract

Diverse human groups produce diverse ideas, the raw material of innovation. Generative AI challenges this engine twice over: everyday AI assistance may homogenize what diverse people create, and AI-simulated diversity may replace the people altogether. We tested both challenges in a preregistered creative metaphor experiment with native (L1) and non-native (L2) English writers, who wrote without AI, with AI-generated ideas (AI ideation), or with AI refining their own ideas (AI refinement). L2 writers contributed more collective diversity than L1 writers, with native-language ideation showing the most diverse pools. AI ideation compressed collective diversity for everyone and left the L2 advantage undetectable, whereas AI refinement preserved both. We then simulated the entire writer pool using personas built from participants' real backgrounds, three model families, native-language prompting, and elevated sampling temperatures. Every simulated pool fell below every human pool, and pushing models further induced diversity only through degenerate text. However, at the individual level, AI ideation raised writers' ratings, pitting private incentives against the collective good, except when L2 writers used their native language, which benefited both. Human diversity remains a valuable creative resource that current AI cannot simulate or sustain; the design of human-AI collaborative workflows determines whether it survives.

## Introduction

Diverse human composition has long been considered a source of creativity and innovation. In scientific research, teams that are diverse in gender, ethnicity, and location produce more novel and higher-impact work (AlShebli et al., 2018; Freeman & Huang, 2015; Yang et al., 2022). Meta-analytic evidence from organizational research echoes this pattern, linking cultural diversity to team creativity and innovation (Wang et al., 2019). Notably, these benefits do not require face-to-face collaboration. Groups of diverse problem solvers outperform groups of high-ability problem solvers even when members contribute independently (Hong & Page, 2004).

Why does who creates shape what gets created? The benefits of diversity are not fully explained by surface-level attributes such as country or ethnicity. Instead, these attributes are proxies for deep-level diversity in values, perspectives, and cognitive repertoires (van Knippenberg & Schippers, 2007). People who represent a problem differently search different regions of the solution space (Hong & Page, 2004). A group of diverse contributors therefore generates a wider pool of ideas, and this variety supplies the raw material for the novel recombinations that drive breakthrough innovation (Page, 2007; Uzzi et al., 2013). This account implies a causal chain: the diversity of contributors (who create) fuels the collective diversity of their output (how varied the resulting pool of ideas is), which in turn fuels innovation. Collective diversity is the middle link that carries the value of diverse composition into creative and innovative outcomes. We adopt this distinction between human diversity and collective diversity throughout this work.

This causal chain now faces two challenges in the age of AI, each attacking a different link. The first challenge concerns the link between diverse contributors and collectively diverse output: even with diverse humans in the loop, everyday use of tools powered by large language models (LLMs) may homogenize their output. Whether writers received ideas from an LLM (Doshi & Hauser, 2024), brainstormed with one (Anderson et al., 2024), or co-wrote with one (Padmakumar & He, 2024), the outputs of different users became more similar to one another. This homogenization has been theorized as a general pressure on human expression and thought (Sourati et al., 2025). The concern is sharpened by incentives. AI assistance reliably boosts individual creative performance and productivity (Doshi & Hauser, 2024; Lee & Chung, 2024; Noy & Zhang, 2023), so individuals are rewarded for adopting the exact tools that homogenize collective output. Homogenization, however, may not be inevitable. For example, introducing an LLM only after independent ideation preserves more original ideas than starting with it (Qin et al., 2025; Zhong et al., 2026). Yet, this literature largely treats users as an undifferentiated pool; little is known about whether LLM assistance erodes the collective benefits of human diversity, or whether those benefits can be protected by how AI enters the workflow.

The second, more radical challenge targets the contributors themselves: human diversity might be simulated. A growing body of “silicon sampling” research shows that LLMs can reproduce human-like judgments and decisions in standard psychological and cognitive tasks (Binz & Schulz, 2023; Dillion et al., 2023). With persona engineering, LLMs can also mimic the responses of diverse subpopulations (Argyle et al., 2023; Park et al., 2024). If simulated personas can stand in for diverse humans, the case for recruiting diverse contributors and building diverse teams weakens. However, this line of research has mostly examined individual-level accuracy against standard benchmarks, not collective-level variance. Where variance has been examined, LLM surrogates respond more uniformly than actual humans (Bisbee et al., 2024; Park et al., 2024), partly because they flatten the heterogeneity of identity groups (Santurkar et al., 2023; Wang et al., 2025). It is less known whether simulated diversity can reproduce the collective diversity of real human groups in open-ended creative production.

Therefore, the current research examines the value of human diversity in creative production in the age of AI. We ask two main questions. First, do diverse humans still confer a collective diversity advantage when they create with LLM assistance? Building on this question, we also explore how such an advantage can be better preserved in AI-assisted workflows. Second, can LLM-simulated diversity substitute for human diversity altogether?

We operationalize human diversity as cultural-linguistic background. Native English (L1) and non-native English (L2) users generated creative metaphors in English. This operationalization rests on two grounds. First, linguistic diversity maps onto deep-level cognitive diversity. Languages differ in how they encode concepts and conventional metaphors (Boroditsky, 2001; Thompson et al., 2020), and these differences shape how people reason (Thibodeau & Boroditsky, 2011). Multilingual experience is itself associated with distinct creative profiles (Kharkhurin, 2010; Leung et al., 2008), and the dominance of English in global knowledge work risks discarding this cognitive variety (Blasi et al., 2022). Second, linguistic diversity captures the domain where AI most credibly helps diverse groups: removing language barriers. Language barriers slow international knowledge diffusion (Higham & Nagaoka, 2025), machine translation measurably expands cross-border exchange (Brynjolfsson et al., 2019), and LLM assistance narrows gaps in writing and creative expression between native and non-native writers (Mei et al., 2025; Shin et al., 2025). For L2 creators, AI is thus both the most promising equalizer and a potential homogenizer, which makes their output the sharpest test case for our questions. As for the creative task, we chose metaphor generation, where such cross-linguistic conceptual differences are most likely to surface. It is a well-validated creative task that recruits both divergent (novelty) and convergent (aptness) processes, and can be scored reliably at scale (DiStefano et al., 2025; Silvia & Beaty, 2012).

To answer the research questions, we employed the following experimental design (**Fig. 1**). The human experiment, including the three conditions and the L1/L2 comparison, was preregistered (link: [https://tinyurl.com/DiverseCreativity](https://tinyurl.com/DiverseCreativity)). First, to vary how much of each idea originated from humans, participants generated metaphors in one of three conditions: a human-only condition (no AI assistance), an AI ideation condition (AI-generated suggestions for both the metaphor and a short explanation), and an AI refinement condition (AI-generated suggestions for the explanation only, after participants came up with their own metaphor). Second, to reflect realistic AI-assisted workflows across language backgrounds, L2 participants in the two AI conditions could choose their preferred language of interaction with AI. Those who worked in a non-English language received AI-generated English translations of their answers before final submission, so all metaphors entered the pool in English. Finally, going beyond the preregistered design, we compared these human pools against LLM-simulated answers generated from personas built on the real participants' background information. We deliberately stacked every analytic choice in the simulation's favor, strengthening the simulation's diversity in three ways: (a) model choice, spanning three families that vary in capability and linguistic coverage (GPT-4o mini; the stronger Claude Sonnet 4.6; and Qwen3.5-27b, trained on more multilingual data); (b) sampling temperature, swept from 0.5 to 2.0, which yields increasingly varied responses; and (c) linguistically diverse prompting, in which simulated L2 personas generated answers in their native languages before translating them into English, with the mix of languages matched to the actual composition of L2 users in the human pool.

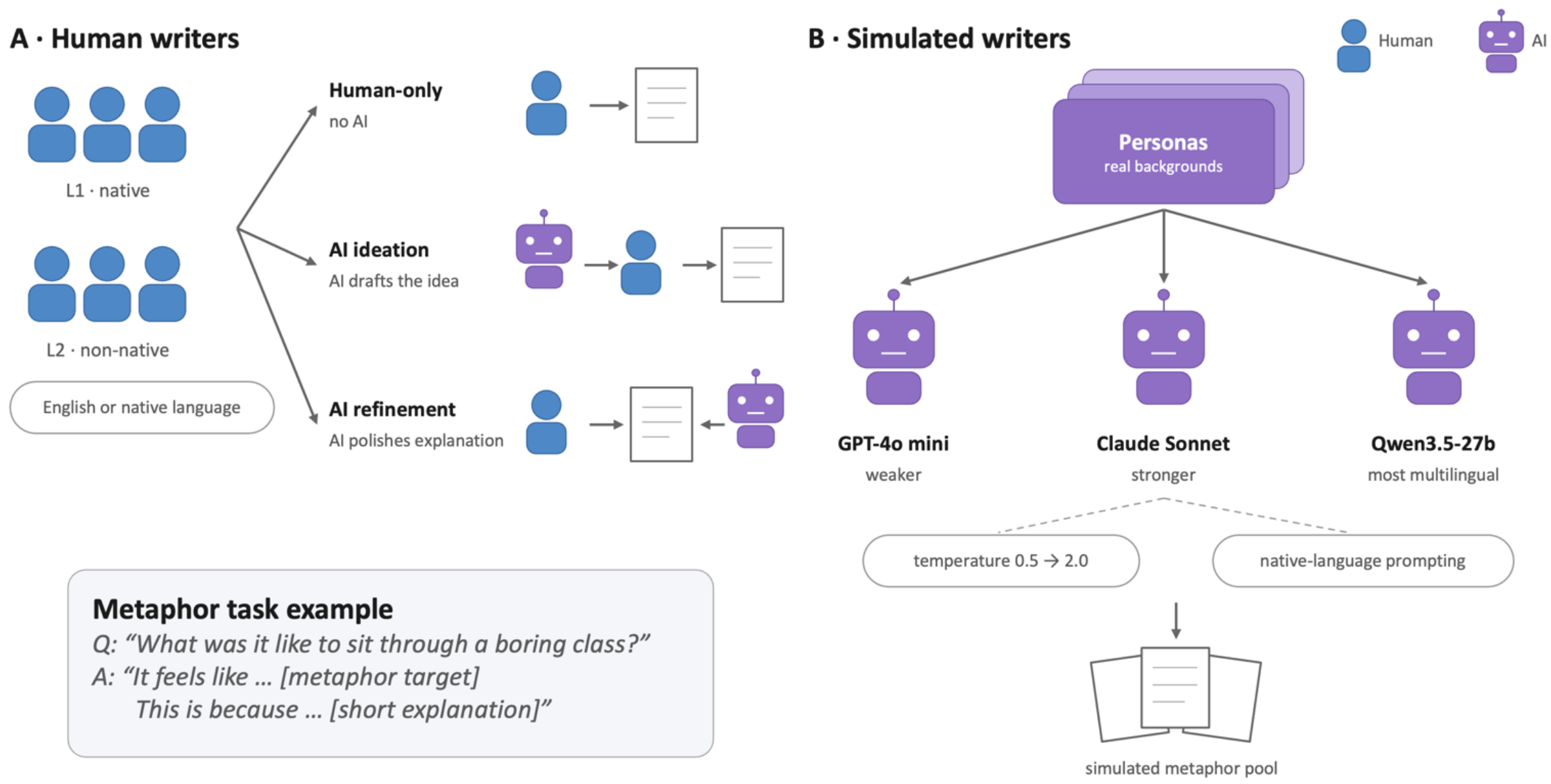

**Fig. 1. Study design. (A)** Native (L1) and non-native (L2) English writers completed four creative metaphor tasks either without AI, with AI-generated ideas (AI ideation), or with AI polishing their own ideas (AI refinement); L2 writers chose their preferred language of AI interaction. **(B)** The same tasks were simulated by LLM personas built from the real participants' backgrounds, with simulated diversity strengthened through model choice, native-language prompting, and higher sampling temperature. The resulting human and simulated metaphor pools were compared on collective diversity.

Results show that, among the human conditions, using AI to ideate reduced collective diversity for both L1 and L2 creative writers, whereas diversity was largely preserved when writers used AI only to refine and explain their self-generated ideas. AI ideation compressed collective diversity and erased the collective advantage of L2 writers, whereas AI refinement preserved both; the advantage was greatest when L2 writers interacted with AI in their native language. Importantly, diverse human groups surpassed LLM simulations in collective diversity, regardless of how we attempted to diversify LLM responses. Finally, analyses of individual creativity suggest why homogenizing workflows may nonetheless spread: AI ideation lifted individual creativity ratings, and these gains depended on AI supplying ideas and were unevenly distributed across L1 and L2 writers. Despite the prevalent, threatening narratives of AI replacing humans (Cave & Dihal, 2019; Frey & Osborne, 2017), our findings highlight that unique and diverse human creative roots remain irreplaceable. The next frontier is to preserve and maximize their value by designing the right human-AI collaboration workflows and by incentivizing authentic, original human contributions.

## Results

Throughout the Results section, we used multilevel regression models to estimate the effects of writing condition and language background (and, among L2 writers, the language of AI interaction) on our outcome measures. We report the key findings in the text; full model specifications and coefficients are in SI Appendix, Tables S1-S3. Below, we first examine whether diverse humans still confer a collective diversity advantage when they write with AI assistance, and how this advantage depends on where AI enters the workflow and on the language of interaction (research question 1). We then test whether LLM-simulated diversity can substitute for the human version (research question 2). Finally, we turn to the individual level, using the creativity ratings from third-party human evaluators as both a quality audit of the human pools and as indicators of the private incentives that determine which workflow writers would be motivated to choose.

**Human linguistic diversity fuels collective diversity, and AI ideation erodes it.** As preregistered, we first examined how AI assistance reshapes the collective diversity of human metaphor pools (**Fig. 2A**). For each condition, we pooled all metaphors written for the same topic and measured how widely they spread in embedding space (mean centroid similarity; **Methods**). If LLM assistance homogenizes creative output, pools with more AI involvement should contract. They did, but the contraction depended on where AI entered the workflow rather than on AI use per se. Writers who received AI-generated ideas produced the most homogeneous pool of the three conditions (vs. human-only: b = 0.036, p < 0.001). Writers who used AI only to refine and explain their own metaphors lost little, remaining statistically indistinguishable from writers without any AI (vs. human-only: b = 0.007, p = 0.516). The collective cost of AI assistance is therefore concentrated at the ideation stage. When AI supplies the ideas, the pool contracts; when humans supply the ideas and AI polishes their expression, the pool survives.

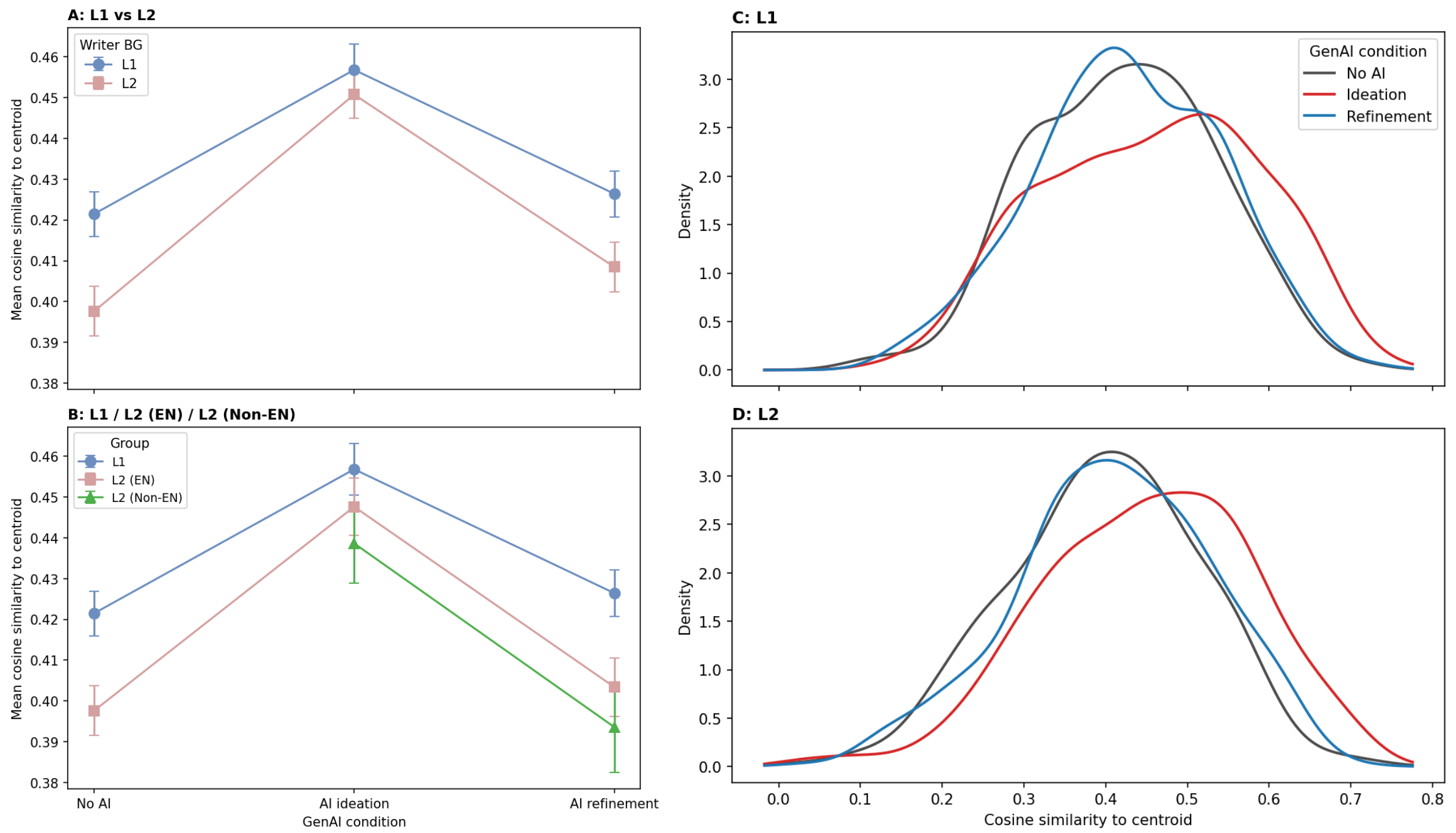

**Fig. 2. Collective diversity of the metaphor embeddings.** For each metaphor, we computed its cosine similarity to the centroid of all other metaphors in the same condition and scenario; higher mean similarity indicates less collective diversity. **(A)** Mean centroid similarity (± SEM) across the three conditions (human-only, AI ideation, AI refinement) for native (L1) and non-native (L2) English writers. **(B)** The same, with L2 split by language of AI interaction, L2 (EN) vs. L2 (Non-EN), the latter present only in the AI conditions. **(C, D)** Kernel density of the per-metaphor centroid similarity (further right = more similar = less diverse), by condition, for **(C)** L1 and **(D)** L2 writers.

We then asked whether the collective advantage of diverse human composition survives these workflows. Without AI, L2 writers produced a more diverse pool than L1 writers (b = -0.022, p = 0.034; L1 vs. L2: M = 0.421 vs. 0.398, Δ = 0.024, p = 0.003), consistent with the premise that linguistic diversity fuels collective diversity. In contrast, AI ideation compressed both groups' pools and narrowed the gap between them, which was no longer statistically distinguishable (L1 vs. L2: $M = 0.457$ vs. $0.451$, $\Delta = 0.006$, $p = 0.481$), although the interaction term itself did not reach significance (language background by writing condition: $b = 0.015$, $p = 0.284$). Under AI refinement, both groups retained near-baseline diversity, and the gap remained with a clear L2 writers' diversity advantage (L1 vs. L2: $M = 0.426$ vs. $0.409$, $\Delta = 0.018$, $p = 0.032$). The collective dividend of human diversity thus survives AI assistance, but only in workflows where humans keep generating the ideas. When AI supplies the ideas, the pool loses breadth, and the advantage of its most diverse members becomes undetectable.

L2 writers who worked in their own language pointed consistently toward greater collective diversity, though the evidence here is directional rather than definitive. This language-choice comparison was preregistered as exploratory, and we interpret it accordingly. In the two AI conditions, L2 writers could interact with the AI in English or in their native language, and 84 of the 228 L2 writers (37%; 39% in AI ideation and 34% in AI refinement) chose the latter. In both conditions, the pools stratified in the same order (**Fig. 2B**). In the mixed model restricted to the AI conditions, the linear contrast across the three ordered groups (L1, L2 EN, and L2 Non-EN writers) was significant ($b = -0.025$, $p = 0.015$). Collective diversity increased monotonically from L1 through L2-English to L2-native writers (centroid similarity, lower = more diverse; $M = 0.442$ vs. $0.427$ vs. $0.421$). Therefore, allowing or even encouraging L2 writers to use AI assistance in their native language could be a simple, low-cost lever for preserving collective diversity, and one that AI translation itself makes practical. A plausible reason is that ideating in one's native language allows writers to search their own conceptual space, drawing on language-specific conventional metaphors and culturally grounded associations before translation flattens them into a shared English idiom.

These results are based on the metaphors themselves. When we instead embedded metaphors together with their explanations (SI Appendix, Fig. S1), the refinement pools also drifted toward the centroid relative to the human-only baseline (AI ideation: $b = 0.028$, $p = 0.019$; AI refinement: $b = 0.046$, $p < 0.001$), and the two AI conditions did not differ ($b = 0.018$, $p = 0.12$). Because this drift appeared only when AI-polished explanations were included, it suggests that AI refinement leaves the ideas diverse but pushes their expression toward a common style. Consistent with this, the diversity advantage of native-language interaction was concentrated in the ideas themselves: once the AI-polished explanations were folded in, it remained directional but no longer significant (L1 vs. L2 Non-EN: $b = 0.017$, $p = 0.14$; L2 EN vs. L2 Non-EN: $b =$

0.020, $p = 0.09$). In short, AI refinement homogenizes how ideas are expressed, whereas the diversity that native-language interaction preserves lies in the ideas themselves.

**LLM-simulated diversity falls below every human pool.** If LLM personas could reproduce this collective diversity, the case for recruiting diverse humans would weaken. We therefore simulated the full writer pool. For every participant in all three human conditions, we built a persona from their background information (including age, sex, ethnicity, education level, first language, etc.; see Fig. S2 for a sample prompt) and asked three model families (GPT-4o mini, Claude Sonnet 4.6, and Qwen3.5-27b) to complete the same set of metaphor tasks. These simulation analyses were not preregistered; we treat them as an adversarial extension and stacked every analytic choice in the simulation's favor. Even so, every simulated pool was less diverse than every human pool in the three conditions (**Fig. 3A**), including the most homogenized human condition, AI ideation [each simulation vs. human with AI ideation: Claude $\Delta = 0.102$, Qwen $\Delta = 0.111$, GPT-4o mini $\Delta = 0.241$, all $p < 0.001$]. The strongest simulation (Claude Sonnet 4.6) recovered only 75.3% of the diversity of the human-only pool. Personas with the same backgrounds as the human writers did not exhibit their variety.

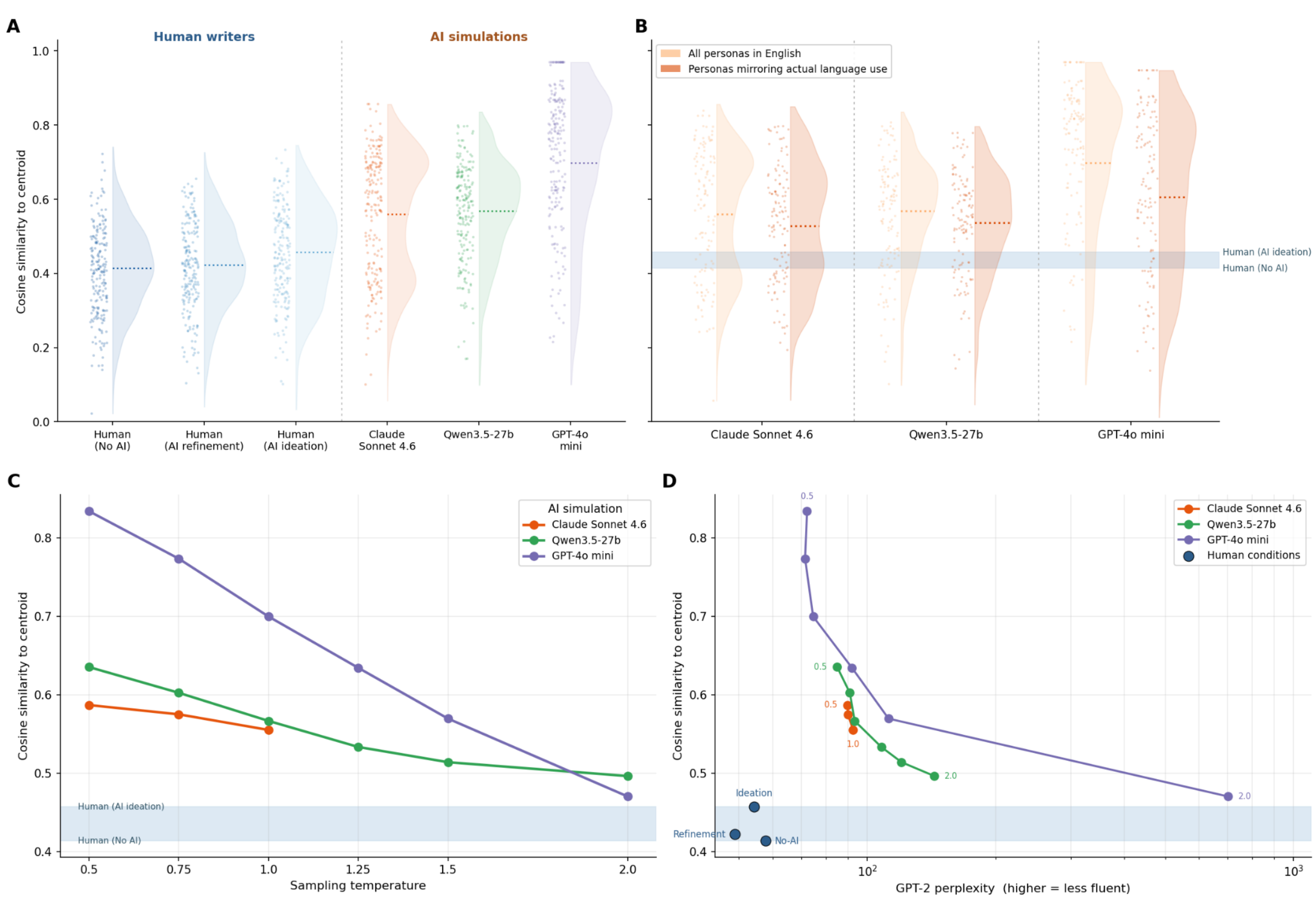

**Fig. 3. Collective diversity of the simulated metaphor pools across models, output language modes, and sampling temperature. (A)** Every LLM simulation (all writers, English output) is more homogeneous than every human condition (all $p < 0.001$; dotted line = mean). **(B)** Generating in each participant's actual language (native for L2 writers who used it, English otherwise) rather than all-English raises diversity, most for the weakest model (GPT-4o mini: $\Delta = 0.09$), but no pool reaches the human range. **(C)** Raising sampling temperature reduces homogeneity, yet diversity saturates above the human range. **(D)** The diversity-fluency frontier (measured by GPT-2 perplexity) places humans as diverse and fluent (lower-left), whereas simulations approach human diversity only by degrading to higher perplexity.

We next granted the simulation the lever that boosted human collective diversity. Simulated L2 personas ideated in their native languages and then translated their answers into English, with the proportion of languages used matched to that of the human pool. This raised simulated diversity across all three models (**Fig. 3B**; see Fig. S3 for subgroup distributions), most visibly for the weakest model (GPT-4o mini including L2 persona with native languages vs. only with English output: $M = 0.606$ vs. $0.698$, $\Delta = -0.092$, $p < 0.001$), mirroring the human native-language effect in direction. But it closed none of the gaps: even the best native-language simulation remained less diverse than the most homogenized human pool in the AI ideation condition (Claude Sonnet 4.6 vs. human with AI ideation: $M = 0.528$ vs. $0.457$, $\Delta = 0.071$, $p < 0.001$). The simulation can imitate the winning workflow; it does not thereby imitate the win.

Sampling temperature was the last and, in principle, the strongest lever, since higher temperature mechanically spreads a model's output distribution. Diversity indeed rose with temperature, but it plateaued around 0.50, leaving a persistent gap of 0.04 to the least diverse human pool (**Fig. 3C**). A single configuration, GPT-4o mini at temperature 2.0, climbed past this plateau, but it did so by breaking: 14.3% of its output words were gibberish (i.e., roughly one in seven words is a non-dictionary string), and format adherence fell below 75%. Plotting collective diversity against fluency (GPT-2 perplexity) makes the constraint visible (**Fig. 3D**). Current models move along a frontier where diversity is bought with fluency. Human pools sit beyond that frontier, jointly diverse and fluent, in a region no simulated configuration has reached. Simulated diversity does not substitute for human diversity; at best, it counterfeits it at the cost of coherence.

**The collectively optimal workflow conflicts with individual incentives.** The collective results identify an optimum: humans generate the ideas, in their native language where applicable, and AI refines the expression. Whether individual writers would choose this arrangement is a different question, because their private returns need not track the collective good. We recruited an independent group of evaluators (N = 351) to rate the metaphors on four items related to creativity: novelty, surprise, metaphor aptness, and explanation quality. Novelty and surprise captured the divergent aspect of creativity (Spearman-Brown $r_{SB}$ = 0.86), the generation of original and unexpected ideas, whereas aptness and explanation quality captured the convergent aspect (Spearman-Brown $r_{SB}$ = 0.87), the production of ideas that are appropriate and coherent for the task. Evaluators only assessed the creative quality of the metaphors and explanations without knowing whether they were produced with AI assistance.

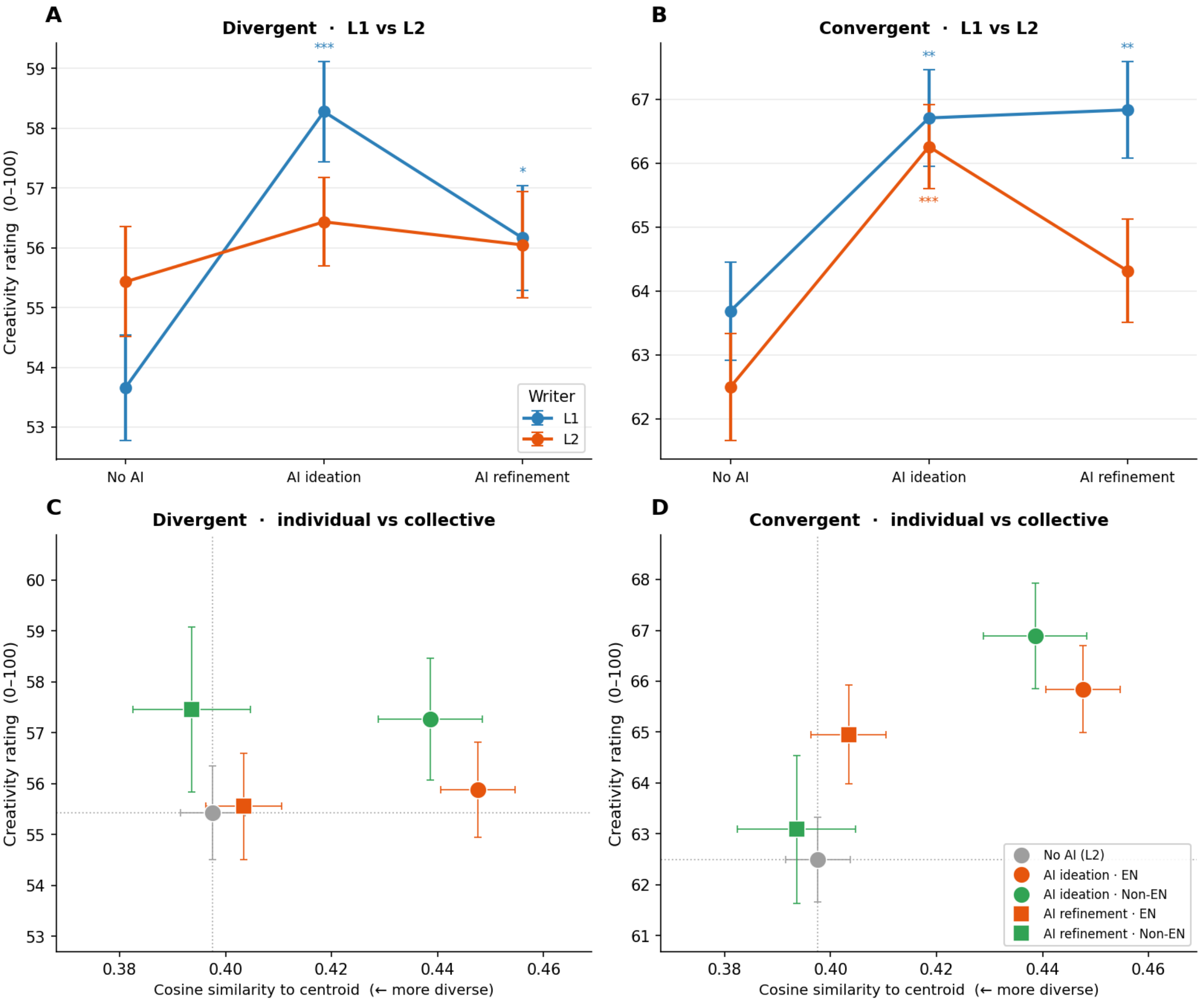

**Fig 4. Individual creativity and its trade-off with collective diversity.** An independent panel of evaluators (N = 351) rated each metaphor on four items, blind to condition; novelty and surprise form the divergent subscale, and aptness and explanation quality form the convergent subscale. **(A, B)** Mean divergent **(A)** and convergent **(B)** rating (± SEM) for native (L1) and non-native (L2) writers. The asterisks indicate the significance level of paired contrasts with the human-only condition within each language group (* $p < .05$, ** $p < .01$, *** $p < .001$). **(C, D)** Incentive landscape for L2 writers: individual divergent **(C)** and convergent **(D)** rating against the collective diversity of the corresponding pool (x-axis, mean

similarity to centroid; lower = more diverse, as in Fig. 2); error bars are ± SEM, and the grey dotted crosshair marks the human-only baseline. Moving up indicates a larger private gain, and moving left indicates greater collective diversity.

Before turning to incentives, the ratings serve a prior purpose: a quality audit of the human pools. We showed above that simulated pools purchase diversity only by degrading fluency. The same suspicion can be raised against human pools, whose collective diversity would mean little if it came from weak or incoherent metaphors, or from writers whose work was judged inferior. Neither was the case. Without AI, L2 writers' metaphors matched L1 writers' on both divergent and convergent creativity (divergent: $b = 2.439$, $p = 0.092$; convergent: $b = -1.221$, $p = 0.306$), suggesting that the group contributing the most collective diversity produced metaphors of comparable individual quality. Similarly, in the diversity-preserving AI workflow, AI refinement improved L1 writers' ratings on both dimensions (divergent: $b = 2.930$, $p = 0.039$; convergent: $b = 3.241$, $p = 0.006$), and left L2 writers' ratings unchanged (divergent: $b = -3.138$, $p = 0.124$; convergent: $b = -1.756$, $p = 0.296$). Together with the fluency results above, this completes the picture: human pools are not only more diverse than simulated ones but also diverse at full quality and full fluency.

The same ratings can also be read as indicators of individual benefit, which exposes an incentive problem: the workflow that pays the writers most is the one that costs the pool most. AI ideation did raise individual ratings, but much of the gain was borrowed rather than owned. However, the gains were asymmetric. AI ideation improved convergent creativity (metaphor aptness and explanation quality) for both groups (L1: $b = 3.127$, $p = 0.007$; L2: $b = 3.711$, $p = 0.001$; **Fig. 4B**). Divergent creativity (novelty and surprise), the dimension closest to collective diversity, rose only for L1 writers ($b = 4.753$, $p < 0.001$) and not for L2 writers ($b = 0.834$, $p = 0.548$; **Fig. 4A**). In the refinement condition, where writers had to produce their own ideas again, these gains receded; L2 writers' convergent advantage was no longer significant ($b = 1.485$, $p = 0.219$; **Fig 4B**) and L1 writers' divergent gain roughly halved ($b = 2.930$, $p = 0.039$; **Fig 4A**). The boost that AI ideation provides to individual performance does not necessarily carry over when the ideas must come from the writer again.

Importantly, the average gains may conceal how they are distributed, and the distribution is where the quality audit becomes consequential. Because innovation draws disproportionately on the best ideas rather than the average idea (Girotra et al., 2010; Wang et al., 2026), how AI reshapes the top of the distribution is a quality question in its own right, and it is also where private incentives are decided. A workflow can raise the mean of a pool while narrowing it, and this is what AI ideation did to L2 writers. Quantile regressions (SI Appendix, Fig. S4) show that AI ideation raised convergent creativity across the rating distribution for both groups, whereas its divergent effect was concentrated in the bottom quantiles for L1 writers ($\beta = +7.5$, $p < 0.001$ at $\tau = 0.1$ vs. $\beta = +4.5$, $p = 0.003$ at $\tau = 0.9$) and existed only there for L2 writers (significant only at $\tau \leq 0.2$). In quality terms, AI ideation raises the floor of the pool but not its ceiling. The most creative metaphors, the ones innovation draws on, gained little or nothing. And these floor-level returns come from precisely the workflow that pulled every pool toward its centroid (**Fig. 2A**). AI refinement offers L1 writers a remedy: it retained their convergent gains (significant at five of nine quantiles, $\tau = 0.2$-$0.7$) and their divergent floor ($\beta = +6.3$, $p = 0.002$ at $\tau = 0.1$ and $\beta = +5.1$, $p = 0.007$ at $\tau = 0.2$) while maintaining the collective diversity. For L2 writers, however, individual gain evaporated under refinement (except for convergent creativity at $\tau = 0.9$; $\beta = +2.6$, $p = 0.048$). The collectively safe workflow is thus backed by only limited private incentives, and, for L2 writers, almost none.

Notably, decomposing L2 writers by their language of AI interaction reveals that this trade-off is not fixed (**Fig. 4C-D**). Measured against the human-only baseline, every AI-assisted cell bought its rating gains with a less diverse pool, except native-language refinement, which was the only cell in the upper-left region of the individual–collective planes. For L2 writers, whether AI assistance pits private benefit against the collective good is thus decided less by whether they use AI than by the language in which they use it.

## Discussion

Human diversity has long been treated as an engine of collective creativity. We asked whether that engine still matters in the age of AI, testing the two challenges it now faces: (1) everyday LLM assistance may erode the collective diversity that diverse humans contribute, and (2) LLM-simulated diversity may replace the humans altogether. Our answer to both is that human diversity retains value that current AI can neither sustain nor simulate by default. Importantly, how AI enters the creative workflow decides how much of that value survives. In our experiment, linguistically diverse (L2) writers contributed more collective diversity than native (L1) English writers, an advantage that AI refinement preserved but AI ideation left undetectable; pools were numerically most diverse when L2 writers ideated in their native language. AI ideation also shrank everyone's pool, whereas AI refinement preserved it. Though optimized in three systemic ways (using varied model families, temperatures, and prompting pipelines), no simulation reached the diversity of any human pool, and pushing models toward more diversity bought it only at the price of degeneration. Blinded ratings of human creative output showed that human diversity carried no cost in rated quality, while revealing a tradeoff between collective diversity and individual benefit. Specifically, the workflow that damages the collective is individually attractive on average. Yet, its private gains are contingent on AI supplying the ideas and thin toward the top of the distribution, barely reaching the most creative L2 writers. And for L2 writers, using their native language to ideate offers a promising solution to align individual and collective interests.

These findings draw a boundary around silicon sampling. Prior work established that persona-conditioned LLMs can approximate the central tendencies of human subpopulations (Argyle et al., 2023; Park et al., 2024) while flattening their heterogeneity (Bisbee et al., 2024; Santurkar et al., 2023; Wang et al., 2025). Our results extend this critique along two dimensions at once: from opinion surveys to open-ended creative production, and from individual-level accuracy to collective-level variance. The failure we document is not a score gap that better prompting might close; it is a structural constraint. Across model families, temperatures, and prompting pipelines, simulated pools moved along a frontier where diversity is purchased at the expense of fluency. In contrast, human pools sat beyond that frontier, jointly diverse and coherent. A simulation that passes item-level plausibility checks can still fail as a population, and it is populations (not items) that creative and innovative ecosystems draw on.

For the homogenization literature, our contribution is to differentiate the users. Prior studies showed that AI assistance makes an undifferentiated pool of users more similar to one another (Anderson et al., 2024; Doshi & Hauser, 2024; Padmakumar & He, 2024). We show that this cost is not spread equally across workflows or writers: it concentrates at the ideation stage, where even the collective advantage of the very population that makes pools diverse becomes undetectable. At the same time, the workflow contrast offers grounds for optimism. When humans supplied the ideas and AI refined the expression, idea diversity survived nearly intact; only expressive style converged. This pattern provides experimental support, at the collective level, on the complement-versus-substitute distinction in models of AI-assisted cultural evolution (Zhong et al., 2026) and on the timing effect observed at the individual level (Qin et al., 2025). The practical principle is to keep humans upstream. Where AI enters the workflow is not a detail of implementation; it is the difference between a tool that polishes a diverse pool and one that quietly replaces it.

The linguistic operationalization of human diversity carries its own implications. Global creative and knowledge work predominantly operates through English, and the pressure on non-native speakers to think and write in English is often discussed as an individual burden (Blasi et al., 2022; Higham & Nagaoka, 2025). Our results suggest it is also a collective cost: forcing L2 writers into English may discard diversity that their native-language ideation would otherwise contribute to the idea pool. The resolution we tested can be called a translate-late principle. AI translation makes it practical to generate ideas in one's own language and translate at the end, rather than thinking in English from the start. In our data, this choice cost little in convergent quality, benefited L2 writers' divergent creativity individually, and was associated with the most diverse pools. This gradient held across conditions and embeddings: pools grew more diverse from L1 through L2-English to L2-native writers. Organizations, journals, and educational settings that default to English-first workflows are, on this evidence, leaving diversity on the table that AI itself can help recover.

The incentive analyses explain why homogenization may spread even where its collective costs are known. AI ideation is attractive to the average writer, and evolutionary modeling suggests that individually favored, substitute-style AI use can prevail even as it drains the variance groups need (Zhong et al., 2026). Countering that drift is a design problem for collectives, not for individuals. Research on collective problem solving may offer a template: what groups must protect is transient diversity, the pool of independently generated candidate solutions that premature convergence destroys (Smaldino et al., 2024), and structures that limit early convergence, such as partial (not fully) connectivity between subgroups, reliably improve collective outcomes (Derex & Boyd, 2016). In reality, the reliance of many individuals on a single family of models may be considered full connectivity to one very persuasive neighbor. Spreading users across providers also does not restore the lost structure. Different model families converge on strikingly similar responses to open-ended prompts (Jiang et al., 2025), and in our own data, all three families fell short of every human pool. The neighbors, in effect, are all the same neighbor. Workflow rules that reserve ideation for humans, norms that allow contributors with diverse backgrounds to work in their own languages and cultural styles, and incentive schemes that reward original rather than efficient contributions are the organizational analogues of those diversity-preserving structures.

Several limitations bound our claims. First, we studied one creative task of short metaphor production; whether the same pattern holds for other forms of innovation (e.g., longer-term, scientific, or visual creativity) remains untested. Additionally, the human-simulation gap may not generalize to more complex contexts, given the brevity and constrained format of our experimental task. Second, we operationalized human diversity as linguistic background; other dimensions of diversity (e.g., gender, ethnicity, country of origin), and their intersections, may behave differently. Third, our substitution claim is bounded in time: it concerns current models under conditions we deliberately tilted in their favor, and future systems or models trained for specific purposes may narrow the gap (e.g., Binz et al., 2025). Fourth, collective diversity was measured on English text; this choice makes the native-language advantage conservative rather than inflated, since all pools were compared in the same language after translation, but embedding-based measures may still miss forms of diversity that human readers would register.

Fifth, our writers produced ideas independently; interacting teams introduce dynamics, from conformity to inspiration, that pooled individual production cannot capture.

The chain from diverse contributors to diverse idea pools to innovation has quietly organized a century of research on creativity and collective performance. Generative AI now presses on both ends of that chain: it offers to stand in for the contributors, and it reshapes what the contributors produce. Our evidence indicates that diverse humans remain a valuable source of collectively diverse creative material we could measure, and the workflows that preserve human ideation, native-language thinking, and AI refinement are promising structural set-ups to maximize their contributions. The question for the AI age is not whether human diversity still matters; it is whether we will keep designing our intelligent assistants and institutional incentives as if it does.

## Materials and Methods

### Participants

**Writers.** We preregistered a target of 600 writers, 100 per cell in the 2 × 3 design, and ultimately recruited 644 writers via Prolific between March 11 and 15, 2026. Native English writers (L1; n = 318) were eligible if indicating English as their sole native language, were born and residing in a majority-English-speaking country. Non-native writers (L2; n = 326) were eligible if indicating a native language other than English. L2 writers reported 34 distinct native languages, most commonly Arabic (15%), Portuguese (13%), Italian (11%), Polish (10%), and Spanish (9%). Writers were on average 36.1 years old (SD = 12.6; range 18-79), 49% women, and predominantly White (66%; Black 12%, Asian 12%; Mixed 4%, Other 6%), and were compensated £3.5 for a median completion time of 20 minutes. Each writer produced four metaphors, one per scenario.

**Evaluators.** An independent sample of 351 native English evaluators (age M = 42.3, SD = 13.5; 49% women; 73% White, 19% Black, 3% Asian, 3% Mixed, 2% Other) was recruited via Prolific on April 13, 2026. Each evaluator rated 24 metaphors: six per scenario, drawn in equal numbers from the three writing conditions and presented in randomized order. This sample size was intended to yield at least three independent ratings per metaphor. Evaluators were compensated £4.5 for a median completion time of 25 minutes.

### Ethics

The study was approved by the ethics committee of the Max Planck Institute for Human Development [NO. A2026-03]. All participants provided informed consent and were informed that their anonymized research data could be shared for research purposes. All data, materials, and analysis code are openly available at
https://osf.io/amgx9/overview?view_only=46c491afd0ba40fc91f361aa2b53be79.

**Writing task and materials**

The writing task followed a 2 × 3 between-subjects design, crossing writer background (L1 vs. L2) with GenAI usage (human-only vs. AI ideation vs. AI refinement). Within the two AI conditions, L2 writers additionally chose their language for AI interaction (English vs. native language), a comparison preregistered as exploratory.

All writers responded to the same four scenarios, presented in a randomized order: (1) "Think of the most boring high-school or college class you've ever had. What was it like to sit through?"; (2) "Think about the most disgusting thing you ever ate or drank. What was it like to eat or drink it?"; (3) "Think about the worst movie or TV show you have ever seen. What was it like to watch it?"; and (4) "Think of the messiest room that you've ever had to live in. What was it like to live there?". For each scenario, writers produced one creative metaphor in a fixed two-part format, "It feels like ___. This is because ___", comprising a metaphor and a short explanation. Instructions emphasized producing novel, surprising comparisons and avoiding clichés.

**Human-only condition.** Writers generated metaphors entirely on their own, in English, with no GenAI assistance provided or allowed at any stage. To deter covert AI use, the survey embedded JavaScript checks that detected tab or window switching: the first switch triggered a warning, and a second switch terminated the session. 24 sessions were terminated on this basis.

**AI ideation condition.** Before composing each metaphor, writers received five AI-generated candidate metaphors, each with a short explanation, generated by GPT-4o (`gpt-4o-2024-08-06`) with the system message reproduced in SI Appendix, Figs. S5 and S6. Writers could adopt, adapt, or ignore the suggestions, and could request regeneration. The final submission was composed in the same two-part format. The same tab-switching detection was applied to prevent the use of external AI tools, resulting in 23 sessions being terminated.

**AI refinement condition.** Writers first composed their own metaphor for the scenario. They then received five AI-generated candidate explanations for their metaphor, generated by GPT-4o (`gpt-4o-2024-08-06`) with the system message in SI Appendix, Figs. S7 and S8, and finalized their explanation, again free to adopt, adapt, or ignore the suggestions. The metaphor itself remained the writers' own. This condition involved 18 sessions being terminated due to tab switching.

**Language of AI interaction.** In the two AI conditions, L2 writers chose whether to interact with the AI in English or in their native language; 84 of 228 L2 writers (37%) chose their native language. When a non-English language was used, the AI provided English translations of the writer's answers, which the writer could edit before final submission. All final submissions were therefore in English, so all pools were compared in the same language.

### Evaluation task

Evaluators rated each metaphor together with its explanation on four items, each on a 0-100 slider (0 = not at all to 100 = extremely). Evaluators were blind to the writers' background and condition, and no information about AI involvement was provided. To counter scale compression, instructions asked evaluators to use the full range and to assign roughly as many ratings below 50 as above 50. The four items were specified as below: (1) Novelty: How novel or original is this metaphor? A novel metaphor is unexpected and avoids familiar or clichéd comparisons;  (2) Surprise: How surprising is this metaphor? Does it catch you off guard in a meaningful way? (3) Appropriateness: How appropriate or fitting is this metaphor? Does the comparison make sense for the scenario described? (4) Explanation quality: How well does the explanation capture why the metaphor works? Does it help you understand the connection between metaphor and experience? Novelty and surprise were averaged into a divergent-creativity subscale (Spearman-Brown $r_{SB}$ = 0.86), and appropriateness and explanation aptness into a convergent-creativity subscale (Spearman-Brown $r_{SB}$ = 0.87), following the standard distinction between the originality and the appropriateness components of creativity (Doshi & Hauser, 2024; Lee & Chung, 2024; Runco & Jaeger, 2012; Silvia & Beaty, 2012).

### Collective diversity measure

Following Doshi and Hauser (2024), we measured collective diversity in embedding space. Each metaphor was embedded with sentence-transformers (`all-MiniLM-L6-v2`); for the primary measure, only the metaphor text ("It feels like ___") was embedded, and for a secondary measure the metaphor and its explanation were embedded together. For each metaphor, we computed the cosine similarity between its embedding and the centroid of all other metaphors written for the same scenario in the same condition (a leave-one-out centroid). Higher mean centroid similarity indicates a more homogeneous, less collectively diverse pool. Comparing pools within scenario controls for topic-level differences.

**LLM simulation of the writer pool**

For every writer in all three conditions, we constructed a persona prompt from their background questionnaire: age, gender, ethnicity, native language, education, household income, subjective social status (0-10 ladder), political orientation, and AI use and literacy (sample prompt in SI Appendix, Fig. S2). Each persona completed the same four scenarios in the same two-part format. We sampled from three model families spanning capability and linguistic coverage: GPT-4o mini (`gpt-4o-mini-2024-07-18`), Claude Sonnet 4.6, and Qwen 3.5 (`Qwen3.5-27b`), each at the provider's default temperature (1.0) for the main comparison.

We then strengthened the simulation's diversity in two further ways. First, in a linguistically diverse pipeline, personas of L2 writers who had used their native language generated their answers in that language and then translated them into English, matching the language mix of the human pool; all other personas generated in English. Second, we swept the sampling temperature from 0.5 to 2.0 for GPT-4o mini and Qwen 3.5, while Claude Sonnet 4.6 supported temperature only up to 1.0. Because outputs at high temperatures can degrade, we checked every simulated metaphor for format adherence, gibberish (the proportion of out-of-vocabulary strings, based on the Python `wordfreq` library), and fluency (perplexity under GPT-2). These simulation analyses were not preregistered; all design choices were made to favor the simulation.

**Statistical analysis**

We analyzed collective diversity (per-metaphor centroid similarity) and creativity ratings with linear mixed-effects models using `statsmodels` 0.14.6 on Python 3.14. Diversity models included random intercepts for writer and scenario; rating models additionally included a random intercept for evaluator. Categorical predictors were treatment-coded, so lower-order coefficients are simple effects at the reference levels (L1, human-only). The ordered-trend contrast across language groups (L1 → L2-EN → L2-Non-EN) was specified post hoc and estimated in an additive model pooling the two AI conditions. Distributional effects on ratings were estimated with quantile regressions at $\tau$ = 0.1-0.9 (in steps of 0.1), fit to metaphor-level mean ratings separately within each writer background. Pool-level comparisons between human and simulated metaphor pools used two-sided permutation tests (10,000 iterations). All tests were two-sided with $\alpha = 0.05$.

### Preregistration and deviations

The experiment, including the three writing conditions, the L1/L2 comparison, and the collective and individual outcome families, was preregistered (link at https://tinyurl.com/DiverseCreativity). The language-choice comparison among L2 writers was preregistered as exploratory. The LLM simulations and the quantile-regression analyses were not preregistered and are reported as exploratory extensions.

The preregistration specified multiple measures for both collective- and individual-level outcome families. Two analytic choices were made after registration, which we note for transparency. First, among the preregistered collective-diversity measures (centroid similarity, semantic distance, and lexical diversity), we report centroid similarity as the primary measure, for comparability with prior work (Doshi & Hauser, 2024). The other preregistered measures are also reported in the SI Appendix, Fig. S9. Second, among the preregistered measures of individual creative performance, we rely on blinded human ratings rather than automated LLM-based scoring. In validation, the LLM scorers produced inconsistent rankings across scoring models and systematically favored LLM-generated content, a circularity that would bias precisely the comparisons this study is about (details in SI Appendix, Fig. S10).

## SI Appendix for

## Human diversity fuels collective creativity that large language models cannot simulate or sustain

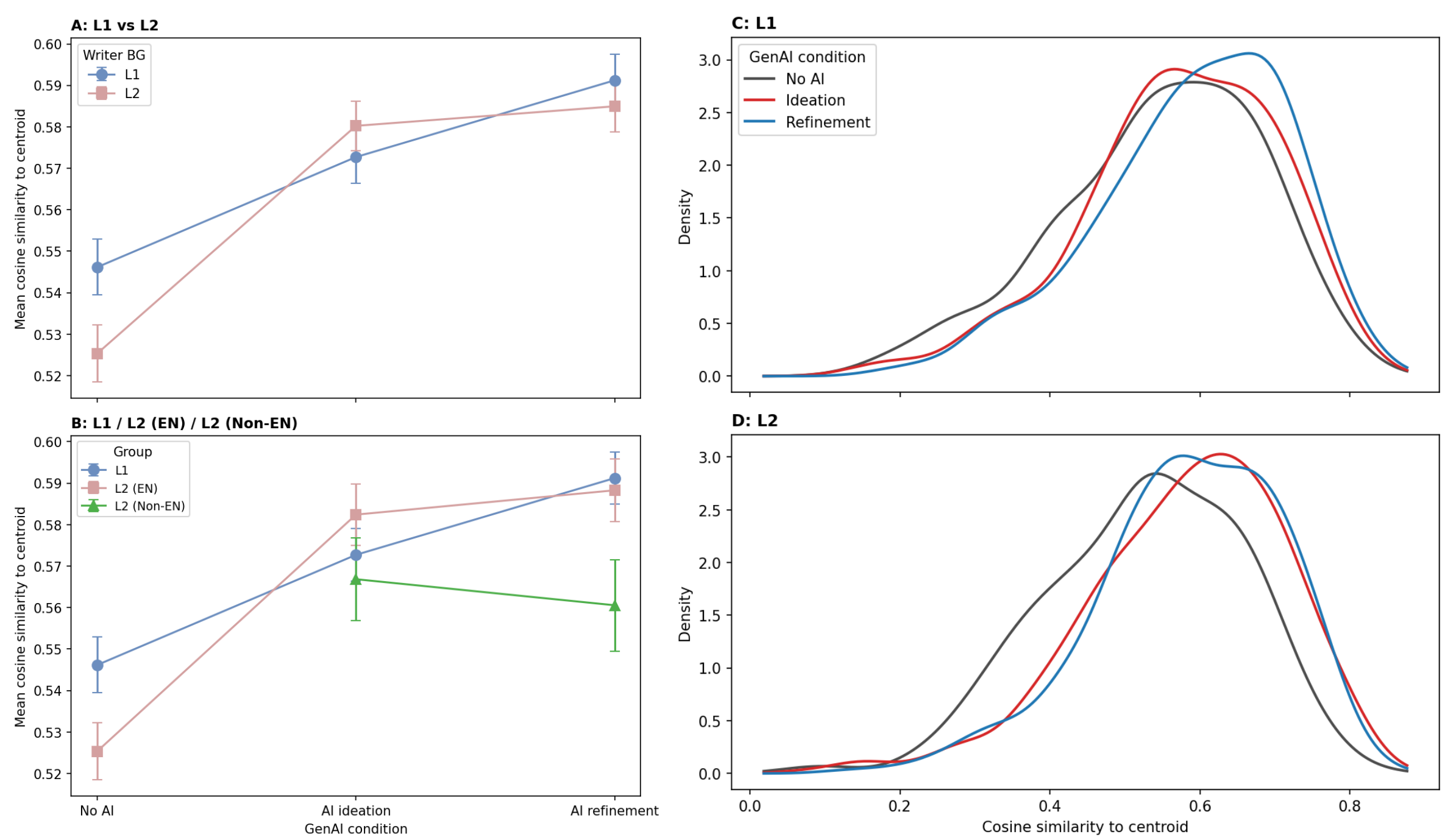

**Fig. S1. Collective diversity of the metaphor + reason embeddings.** As in Fig. 2, but each metaphor is embedded together with its explanation. For each metaphor, we computed its cosine similarity to the centroid of all other metaphors in the same condition and scenario; higher mean similarity indicates less collective diversity. **(A)** Mean centroid similarity (± SEM) across the three conditions for L1 and L2 writers. **(B)** The same, with L2 split into L2 (EN) vs. L2 (Non-EN). **(C, D)** Kernel density of the per-metaphor centroid similarity (further right = more similar = less diverse), by condition, for L1 **(C)** and L2 **(D)** writers.

```
You are a 31-year-old man White person. English is your native language.
Your education: Graduate degree (MA/MSc/MPhil/other). Your household
income is around $30000–$39999. On a 0-10 social ladder of standing in
your community, you place yourself at 4. Your political orientation is 1
on the survey's scale. You use AI tools: weekly. Your AI literacy is
high.

You are taking part in a creativity study. You will be given a scenario
and must respond with a single creative metaphor describing how it feels.
A creative metaphor is novel and surprising — avoid clichés and
predictable comparisons. Answer in English using EXACTLY this one-line
format, filling in the two blanks and writing nothing else:

It feels like ___. This is because ___.

Scenario: Think about the most disgusting thing you ever ate or drank.
What was it like to eat or drink it?
```

**Fig. S2. Example simulation prompt, constructed for a persona of an L1 participant.** The system message concatenates the participant's real background (age, gender, ethnicity, native language, education, household income, social-ladder standing, political orientation, and AI use and literacy) with the task instruction; the user message supplies one of the four scenarios.

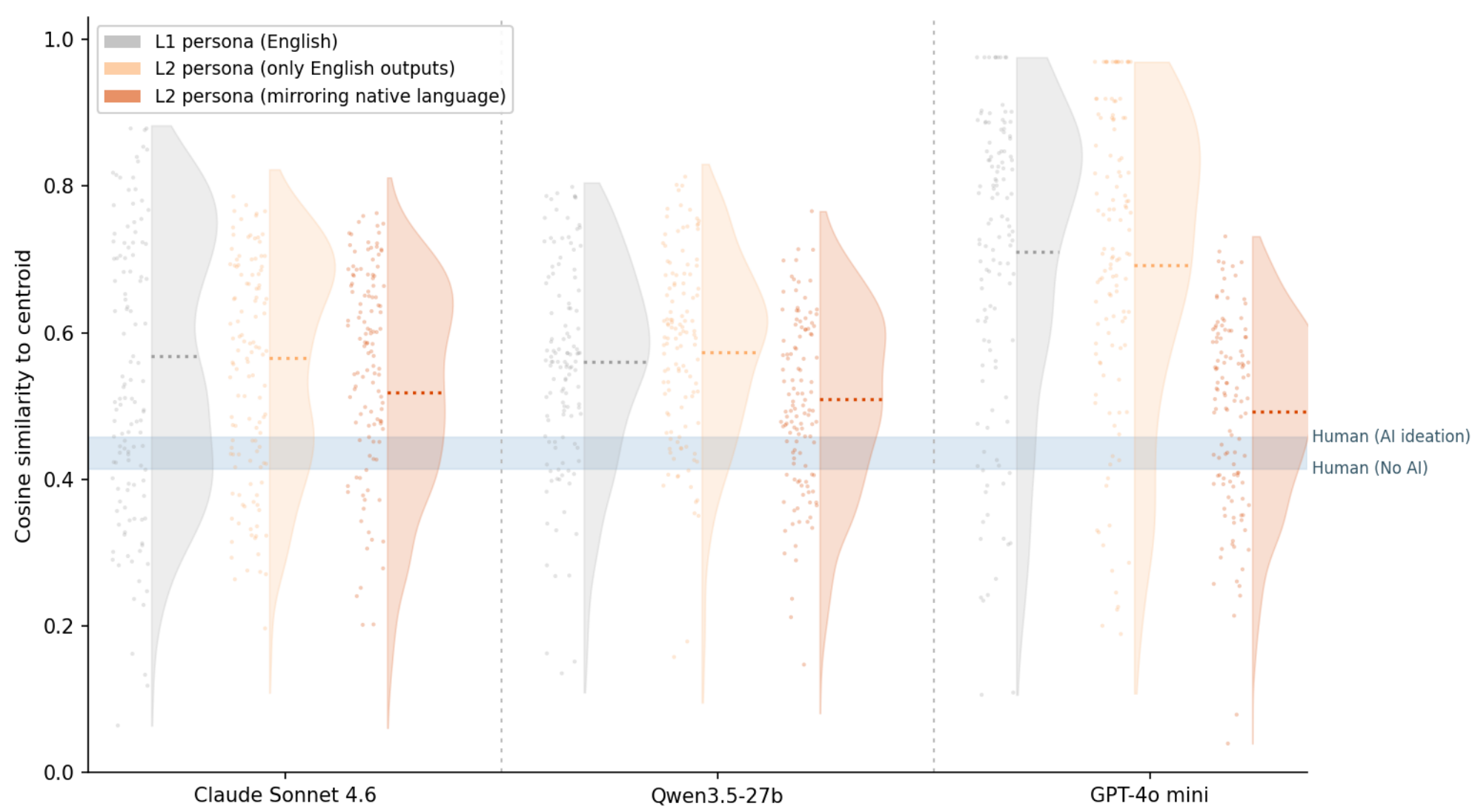

**Fig. S3. Subgroup decomposition of the collective diversity in Fig. 3B by persona background.** Cosine similarity to Centroid (dotted line = mean; blue band = human range) for the Ideation and Refinement personas (i.e., the writers who had the native-language option) split into L1 personas (English), L2 personas (only English outputs), and L2 personas (mirroring native language). The gain in Fig. 3B is concentrated in L2 personas: generating in the native language raises L2 diversity for every model (Claude Δ = 0.05, Qwen Δ = 0.06, GPT-4o mini Δ = 0.20; all $p < 0.001$), whereas L1 personas, which always generate in English, are unchanged. No subgroup reaches the human range.

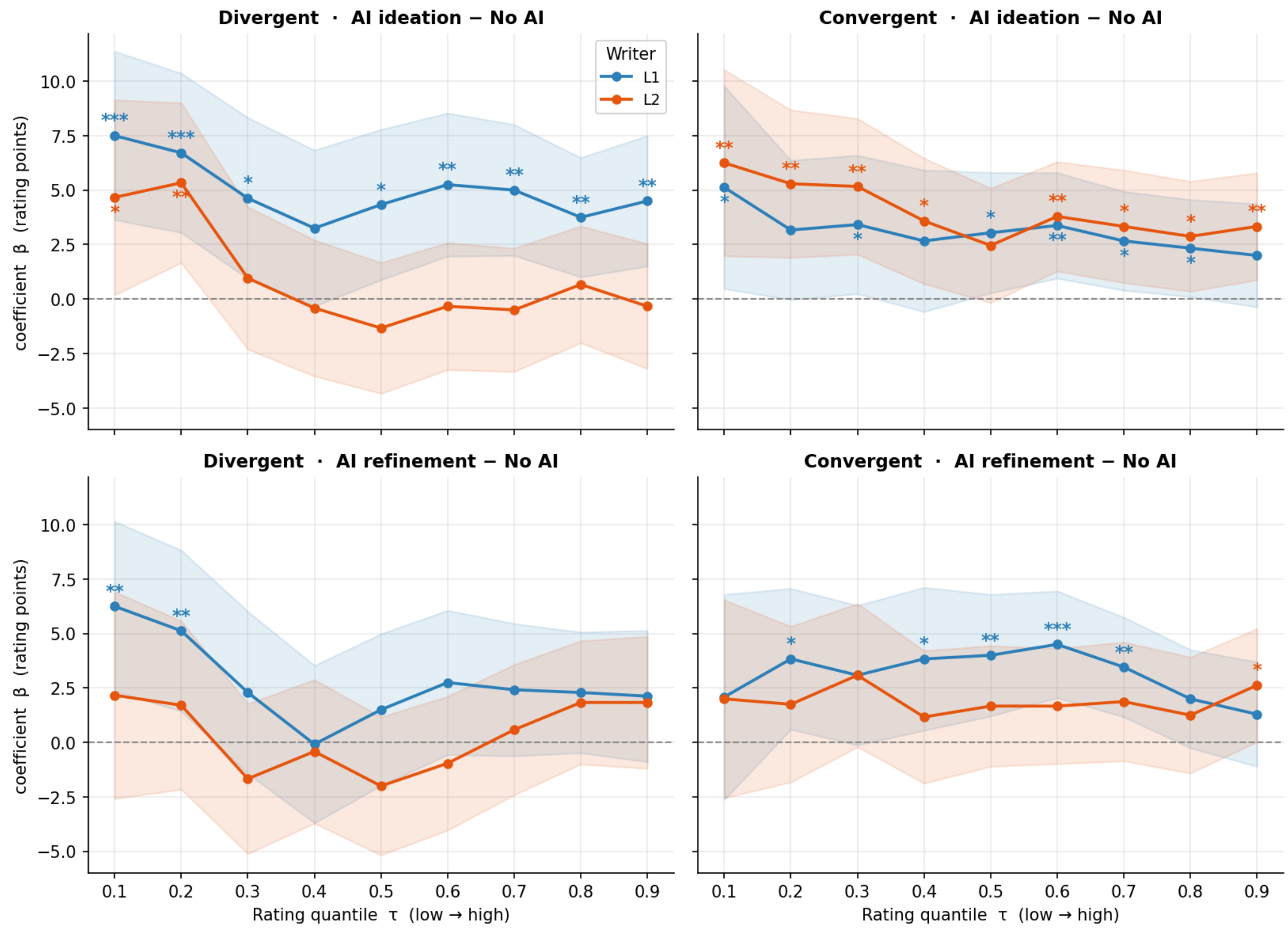

**Fig. S4. Quantile regressions of individual creativity ratings, fit separately for native (L1) and non-native (L2) writers.** Each panel plots the condition coefficient β (points; vs No AI) across rating quantiles τ with its 95% CI (shaded band). Stars mark quantiles at which β differs from zero ($p < 0.05$), placed above the curve where that group is higher than the other and below otherwise. AI ideation raised L1 writers' divergent creativity across the whole distribution but L2 writers' only at the lowest quantiles ($\tau \leq 0.2$), while raising convergent creativity broadly for both groups.

```
The user is tasked with coming up with a metaphor for the following
scenario: "Think of the most boring high-school or college class you've
ever had. What was it like to sit through?" Their aim is to come up with
something creative: something clever, humorous, original, compelling, or
interesting. The answer must be in English and in the format: "It feels
like ______. This is because ______." where the second blank should span
two to three sentences.

Your role is to guide the user in the following steps:
1. Repeat the above scenario and provide 5 suggestion pairs of the first
and second blanks, where the latter of each should span two to three
sentences. (Note that the user could ask for further rounds of
suggestions. Also, the suggestions should be written in first person so
that the user can reuse them.)
2. Continue the conversation until the user is satisfied.
```

**Fig. S5. Example AI ideation prompt, prepared for L1 participants.** The scenario shown is one of the four, and the same template was used for all four.

```
The user is tasked with coming up with a metaphor for the following
scenario: "Think about the most disgusting thing you ever ate or drank.
What was it like to eat or drink it?" Their aim is to come up with
something creative: something clever, humorous, original, compelling, or
interesting. While you have to respond in [specified language] during the
conversation, the final answer must be in English and in the format: "It
feels like ______. This is because ______." where the second blank should
span two to three sentences.

Your role is to guide the user in the following steps:
1. Repeat the above scenario and provide 5 suggestion pairs of the first
and second blanks in [specified language], where the latter of each
should span two to three sentences. (Note that the user could ask for
further rounds of suggestions. Also, the suggestions should be written in
first person so that the user can reuse them.)
2. Continue the conversation until the user is satisfied.
3. Once the user seems satisfied, suggest offering the final English-
translated version of their response for the two blanks.
```

**Fig. S6. Example AI ideation prompt, prepared for L2 participants.** The field [specified language] is automatically filled with the language each participant has indicated they want to use for the interaction.

```
The user is tasked with coming up with a metaphor for the following
scenario: "Think about the worst movie or TV show you have ever seen.
What was it like to watch it?" Their aim is to come up with something
creative: something clever, humorous, original, compelling, or
interesting. The answer must be in English and in the format: "It feels
like ______. This is because ______." where the second blank should span
two to three sentences.

Your role is to guide the user in the following steps:
1. Repeat the above scenario and ask to provide a short answer to the
first blank. (Note that you are prohibited from providing any answers or
hints for the first blank, to avoid biasing the user. You MUST NOT
present any questions or suggestions until the user provides their answer
for the first blank; what you can do is only asking the user to provide
their own answer to the first blank.)
2. With the user's entered answer to "It feels like …", provide 5
suggestions for the second blank "This is because …", where each should
span two to three sentences. (Note that the user could ask for further
rounds of suggestions. Also, the suggestions should be written in first
person so that the user can reuse them.)
3. Continue the conversation until the user is satisfied.
```

**Fig. S7. Example AI refinement prompt, prepared for L1 participants.** The scenario shown is one of the four, and the same template was used for all four.

```
The user is tasked with coming up with a metaphor for the following
scenario: "Think about the most disgusting thing you ever ate or drank.
What was it like to eat or drink it?" Their aim is to come up with
something creative: something clever, humorous, original, compelling, or
interesting. While you have to respond in [specified language] during the
conversation, the final answer must be in English and in the format: "It
```

```
feels like ______. This is because ______." where the second blank should
span two to three sentences.

Your role is to guide the user in the following steps:
1. Repeat the above scenario and provide 5 suggestion pairs of the first
and second blanks in [specified language], where the latter of each
should span two to three sentences. (Note that the user could ask for
further rounds of suggestions. Also, the suggestions should be written in
first person so that the user can reuse them.)
2. Continue the conversation until the user is satisfied.
3. Once the user seems satisfied, suggest offering the final English-
translated version of their response for the two blanks.
```

**Fig. S8. Example AI refinement prompt, prepared for L2 participants.** The field [specified language] is automatically filled with the language each participant has indicated they want to use for the interaction.

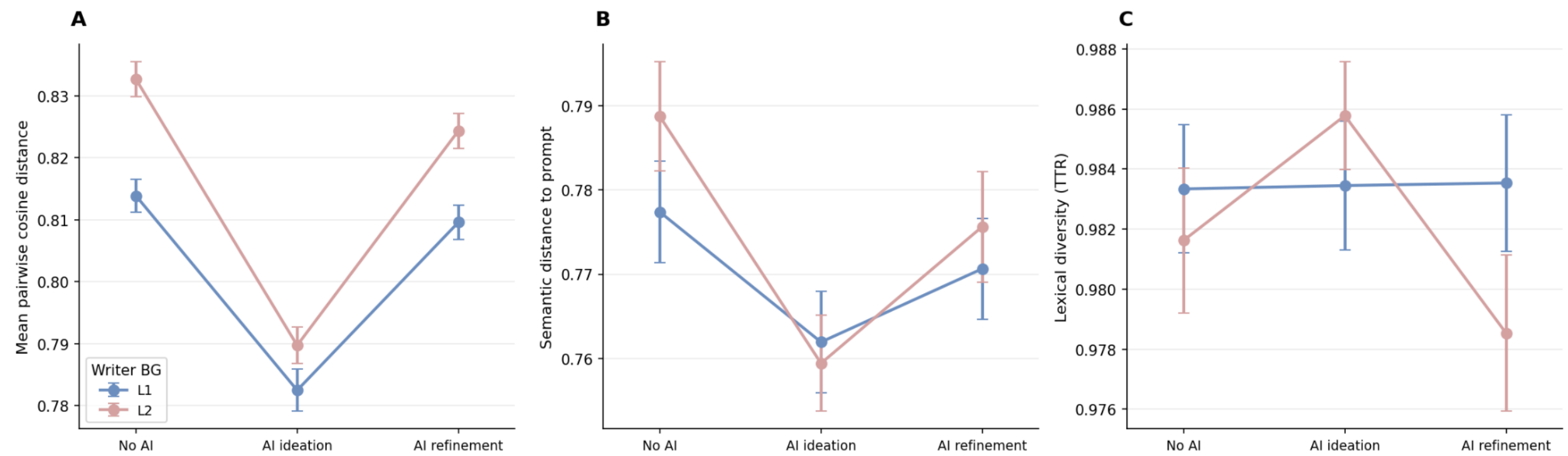

**Fig. S9. Collective diversity under the other measures.** The primary result is unchanged: AI ideation reduces collective diversity for both groups and narrows the L1–L2 gap, whereas AI refinement preserves it. However, lexical diversity (C) is near the ceiling and uninformative for these short metaphors.

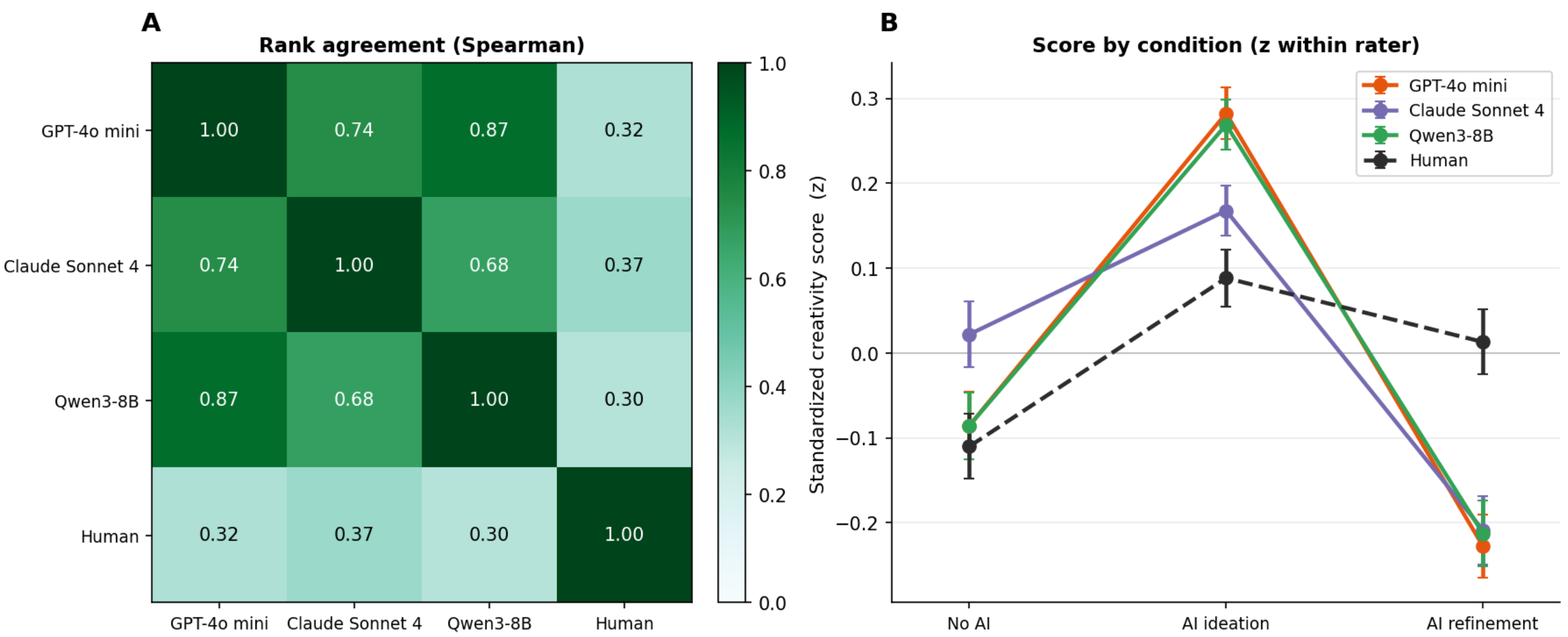

**Fig. S10. LLM-based creativity scoring calibrated based on the Bradley-Terry model.** Three LLM judges (GPT-4o mini, Claude Sonnet 4, Qwen3-8B) repeatedly ranked a set of randomly sampled metaphors; scores are compared with one another and with the blinded human panel (mean rating per metaphor; n = 2,235). (A) Spearman rank agreement: the LLM judges track one another moderately (ρ = 0.68–0.87) but track the human panel poorly (ρ = 0.31–0.38). (B) Standardized mean score by condition

(z within rater; mean ± SEM): two of the three LLM judges (GPT-4o mini, Qwen3-8B) sharply over-reward AI-ideation metaphors and penalize AI-refinement, a condition that the human panel does not produce.

---

**Tab. S1.** diversity ~ writer_bg * condition + (1|writer) + (1|scenario)

| Predictors | Metaphor: *b* | 95% CI | p | Metaphor + reason: *b* | 95% CI | p |
|---|---|---|---|---|---|---|
| (Intercept) | 0.424 | 0.385 – 0.463 | <0.001 | 0.548 | 0.507 – 0.589 | <0.001 |
| AI ideation (vs No-AI) | 0.036 | 0.017 – 0.056 | <0.001 | 0.028 | 0.005 – 0.051 | 0.019 |
| AI refinement (vs No-AI) | 0.007 | -0.013 – 0.027 | 0.516 | 0.046 | 0.022 – 0.069 | <0.001 |
| L2 writer (vs L1) | -0.022 | -0.043 – -0.002 | 0.034 | -0.020 | -0.043 – 0.004 | 0.107 |
| AI ideation × L2 | 0.015 | -0.013 – 0.043 | 0.284 | 0.027 | -0.006 – 0.059 | 0.107 |
| AI refinement × L2 | 0.005 | -0.024 – 0.034 | 0.730 | 0.013 | -0.021 – 0.046 | 0.456 |
| **Random effects** | | | | | | |
| $\sigma^2$ | 0.011 | | | 0.011 | | |
| $\tau_{00}$ writer | 0.003 | | | 0.004 | | |
| $\tau_{00}$ scenario | 0.001 | | | 0.001 | | |
| ICC | 0.27 | | | 0.34 | | |
| $N_{writer}$ | 644 | | | 644 | | |
| $N_{scenario}$ | 4 | | | 4 | | |
| Observations | 2493 | | | 2493 | | |
| Marginal $R^2$ / Conditional $R^2$ | 0.028 / 0.293 | | | 0.029 / 0.363 | | |

**Tab. S2.** score ~ writer_bg * condition + (1|evaluator) + (1|writer) + (1|scenario)

| Predictors | Divergent: *b* | 95% CI | p | Convergent: *b* | 95% CI | p |
|---|---|---|---|---|---|---|
| (Intercept) | 53.41 | 50.36 – 56.46 | <0.001 | 63.65 | 61.07 – 66.23 | <0.001 |
| AI ideation (vs No-AI) | 4.75 | 2.00 – 7.50 | <0.001 | 3.13 | 0.86 – 5.39 | 0.007 |
| AI refinement (vs No-AI) | 2.93 | 0.15 – 5.71 | 0.039 | 3.24 | 0.95 – 5.53 | 0.006 |
| L2 writer (vs L1) | 2.44 | -0.40 – 5.27 | 0.092 | -1.22 | -3.56 – 1.11 | 0.306 |
| AI ideation × L2 | -3.92 | -7.79 – -0.05 | 0.047 | 0.58 | -2.61 – 3.78 | 0.720 |
| AI refinement × L2 | -3.14 | -7.13 – 0.86 | 0.124 | -1.76 | -5.05 – 1.54 | 0.296 |
| **Random effects** | | | | | | |
| $\sigma^2$ | 477.68 | | | 410.91 | | |
| $\tau_{00}$ evaluator | 95.20 | | | 108.96 | | |

| | | |
|---|---|---|
| $\tau_{00}$ writer | 66.54 | 38.12 |
| $\tau_{00}$ scenario | 4.63 | 3.01 |
| ICC | 0.26 | 0.27 |
| N $_{evaluator}$ | 351 | 351 |
| N $_{writer}$ | 644 | 644 |
| N $_{scenario}$ | 4 | 4 |
| Observations | 8241 | 8241 |
| Marginal $R^2$ / Conditional $R^2$ | 0.003 / 0.261 | 0.005 / 0.271 |

**Tab. S3.** The same analysis as Tab. S1 and S2, comparing subgroups of L2 against L1.

| Predictors | Divergent rating: b | 95% CI | p | Centroid similarity: b | 95% CI | p |
|---|---|---|---|---|---|---|
| (Intercept) | 57.86 | 55.06 – 60.65 | <0.001 | 0.460 | 0.421 – 0.499 | <0.001 |
| L2-EN (vs L1) | −1.60 | −3.75 – 0.56 | 0.147 | −0.016 | −0.032 – 0.001 | 0.060 |
| L2-Non-EN (vs L1) | 0.03 | −2.54 – 2.60 | 0.983 | −0.025 | −0.044 – −0.005 | 0.015 |
| AI refinement (vs ideation) | −1.31 | −3.21 – 0.60 | 0.178 | −0.037 | −0.052 – −0.023 | <0.001 |
| Random effects | | | | | | |
| $\sigma^2$ | 470.60 | | | 0.011 | | |
| $\tau_{00}$ evaluator | 91.57 | | | — | | |
| $\tau_{00}$ writer | 63.82 | | | 0.003 | | |
| $\tau_{00}$ scenario | 4.27 | | | 0.001 | | |
| ICC | 0.25 | | | 0.30 | | |
| N evaluator | 351 | | | — | | |
| N writer | 441 | | | 440 | | |
| N scenario | 4 | | | 4 | | |
| Observations | 5638 | | | 1707 | | |
| Marginal $R^2$ / Conditional $R^2$ | 0.002 / 0.255 | | | 0.026 / 0.316 | | |